\documentclass[%
 reprint,
 amsmath,amssymb,
 aps,
 prl,
 longbibliography,
 lengthcheck,%
]{revtex4-1}

\usepackage{graphicx}
\usepackage{dcolumn}
\usepackage{bm}
\usepackage{hyperref}

\begin{document}

\preprint{APS/123-QED}

\title{Soliton appearing in boson-fermion mixture at the third order of the interaction radius}

\author{K. V. Zezyulin}
\email{konstantin.zez@mail.ru}

\author{P. A. Andreev}%
\email{andreevpa@physics.msu.ru}
\author{L. S. Kuz'menkov}%
 \email{lsk@phys.msu.ru}
\affiliation{%
 Physics Faculty, Moscow State
University, Moscow, Russian Federation.}%

\begin{abstract}
In this paper we consider an ultra-cold mixture of boson and
fermion atoms on the basis of quantum hydrodynamics. Small
perturbations in such systems are being analyzed. A possibility is
shown for soliton solutions of a new type to appear if the third
order of the interaction radius is taken into account in the
analysis of interactions. A fermion-fermion interaction occurs in
explicit form if this approximation is accepted. The conditions
that lead to occurrence of this type of soliton in a mixture of
boson and fermion atoms were investigated. Restrictions on the
fermion-fermion interaction were found that are necessary for this
kind of perturbations to appear in the system. Conditions
determining whether perturbances would be a condensed soliton or a
rarefied soliton are shown. Requirements to the experimental
detection of a new soliton type in boson-fermion mixture are
considered.
\end{abstract}


\maketitle

\section{\label{sec:level1}I. Introduction}

Experimental and theoretical approaches to the physics of quantum
gases have been continued to grow in recent times, and various
nonlinear structures has been studied there ~\cite{Parker PRA
12}-~\cite{Filatrella PRA 09}. Methods for producing Bose-Einstein
condensate and investigating its properties have been developed
~\cite{Denschlag Science 00}. This showed stable conditions to
exist both for repulsion and attraction between the atoms, for
example, in a Bose condensate of $^7Li$ atoms ~\cite{Strecker
Nature 02}. A mixture of Bose and Fermi degenerate gases, where
interaction between atoms may be tuned from repulsion to
attraction, is considered in ~\cite{Karpiuk PRL 04}. Papers
~\cite{Karpiuk PRL 04, Ramachandhran PRA 11} have been devoted to
investigations of ultra-cold boson-fermion mixtures. A branch of
physics that works on producing of a stable boson-fermion mixture
is being developed as well \cite{Zeng-Qiang Yu PRA 11}.

A lot of interest has been attracted in recent years to the
investigation of soliton solutions. Papers devoted to properties
of different types of solitons are constantly being emerged. As to
the present time a bright soliton has been produced in a
Bose-Einstein condensate with a repulsive interaction between
atoms. A soliton of this type, has been produced, for example, in
the condensate of $^{87}Rb$ atoms in a weak periodic potential
\cite{Eiermann PRL 04}. The formation of a bright soliton in the
condensate of $^7Li$ atoms is considered in ~\cite{Strecker Nature
02, Khaykovich Science 02} for a quasi-1D case. Properties of a
dark soliton have been studied in \cite{Burger PRL 99}. They show
experimental generation of a dark soliton in a cigar-shaped
condensate of $^{87}Rb$ atoms with the use of phase imprinting
technique. The study of a dark soliton in Bose-Einstein condensate
inside a vortex ring is described in \cite{Anderson PRL 01}. In
addition to one-dimension solitonic perturbations in a
Bose-Einstein condensate, three-dimension perturbations are also
attracting attention  \cite{Mateo PRA 11}. A possibility was shown
for solitonic perturbations to exist in a Bose-Einstein condensate
with dipole-dipole interaction \cite{Cuevas PRA 09}. In recent
years predictions have been made that solitons of bright-bright,
dark-bright and dark-dark types may exist in a two-component BEC
\cite{Csire PRA 10}. Solitons in boson-fermion mixture is special
and interesting item. Perturbations in boson-fermion mixtures have
also been studied and the existence of a bright soliton in
boson-fermion mixture has been proved in ~\cite{Karpiuk PRL 04,
Adhikari PRA 07}.

In the present work we study ultra-cold boson-fermion mixtures.
Examples of such mixtures are  $^7Li-^6Li$ \cite{Truscott Science
01, Schreck PRL 02}, $^{23}Na-^6Li$ \cite{Hadzibabic PRL 02} and
$^{87}Rb-^{40}K$ \cite{Roati PRL 02}.The possibility of new
solitonic perturbations in a mixture of boson and fermion atoms
with short-range interaction potentials is analyzed in particular.
Theoretical analysis at higher order of interaction account
accuracy allows prediction of new types of soliton solutions
\cite{Andreev PRA08}. This approximation can be derived with the
aid of quantum hydrodynamics approach, which has been developed in
recent decade \cite{Andreev PRA08,MaksimovTMP 1999,Andreev arxiv
12 02}. The set of quantum hydrodynamic equations describing an
ultra-cold mixture of boson and fermion atoms may be derived from
a many-particle Schrodinger equation \cite{Andreev PRA08}. In
\cite{Andreev RPJ 11}, for example, they used this approach to
study how the shape of a bright soliton changes as interactions in
a Bose-Einstein condensate are considered more precisely. The
method of quantum hydrodynamics is very useful in different areas
of physics. Namely, it was quantum plasma ~\cite{MaksimovTMP
1999}, ~\cite{Haas PRE 00}, plasma of particles with the own
magnetic moment ~\cite{MaksimovTMP 2001}-~\cite{Shukla RMP 11},
relativistic quantum plasma ~\cite{Asenjo PP 11}, ultracold Bose
and Fermi gases with nonlocal interaction ~\cite{Andreev PRA08},
quantum particles with electrical ~\cite{Andreev PRB 11},
~\cite{Andreev arxiv 12 02} and magnetical ~\cite{Andreev arxiv 12
MagnPol} polarization, particularly being in the BEC state
~\cite{Andreev arxiv 12 02}, ~\cite{Andreev arxiv 12 MagnPol}, the
graphene electrons ~\cite{Andreev arxiv 12 01} and BEC of graphene
excitons ~\cite{Andreev arxiv 12 GEBEC}.

Most of theoretical papers devoted to the ultracold fermions
contain nonlinear Schrodinger equation (an analog of
Gross-Pitaevskii equation using for bosons) which does not include
interaction ~\cite{Butts PRA97}-~\cite{Adhikari PRA 04}. It is
connected with the fact that the first Born approximation gives no
contribution in interaction of ultracold fermions. This equation
undoubtedly accounts the Fermi pressure contribution caused by
Pauli principle. It is very interesting to understand a role of
interaction in fermion systems. For example, vortices in a
strongly interacting gas of fermionic atoms on the BEC- and the
BCS-side of the Feshbach resonance was experimentally studied in
Ref. ~\cite{Zwierlein  Nat 05}. Presented in our paper model based
on the quantum hydrodynamics method contain fermion-fermion
interaction whose form was derived directly from the Schrodinger
equation ~\cite{Andreev PRA08}.

In this paper we discuss the existence of new types of solitons in
a mixture of boson and fermion atoms and determine the conditions
of their experimental detection. A substantial role in our
analysis is given to the fermion-fermion interaction coefficient,
which appears in the analysis of the third order of the
interaction range. A perturbation method introduced by Washimi
\cite{Washimi PRL 66}, \cite{Infeld book} is used here to find
solitonic perturbations. An analogous method was used for a
spinor-1 BEC studying ~\cite{Cai CTP 11}. Other methods for
weak-nonlinear analysis of BEC have been considered in literature.
For example, in Ref. ~\cite{Andreev Izv.Vuzov. 09 1} the
Krylow-Bogoliubov-Mitropolskii method was used for nonlinear
frequency shift calculation. It was shown that account of
interaction up to TOIR approximation leads to the new solitons in
BEC ~\cite{Andreev arxiv 11 1}.

Our paper is organized as follows. In Sect. 2 we present basic
equation and describe using model. In Sect. 3 we study solitons in
boson-fermion mixture which appearing due to interaction account
up to TOIR approximation. In Sect. 3 detailed analysis of
conditions of a "fermion" soliton existence is presented. In Sect.
4 condition of a "boson" soliton existence is described. In Sect.
5 brief summary of obtained results is presented.

\section{\label{sec:level1}II. The model}

Let's consider a mixture of ultra-cold boson and fermion atoms
having our focus on one-dimension perturbations. The set of
quantum hydrodynamic equations consists of the following
equations: the continuity equation for bosons
\begin{equation} \label{boson_nepr}
\frac{\partial n_b}{\partial t}+\frac{\partial (n_b v_b
)}{\partial x}=0 ,
\end{equation}
the momentum balance equation for bosons
$$m_b n_b  \frac{\partial v_b}{\partial t}+\frac{1}{2} m_b n_b  \frac{\partial v_b^2}{\partial x}-\frac{\hbar^2}{2m_b } n_b  \frac{\partial}{\partial x} \biggl( \frac{ 1 }{\sqrt{n_b }}\frac{\partial^2 \sqrt{n_b}}{{\partial x}^2}\biggr)$$
$$- \Upsilon_{bb} n_{b} \frac{{\partial n_b}}{\partial x}- \frac{1}{16} \Upsilon_{2bb}  \frac{{\partial}^3 n_b^2}{{\partial x}^3 }$$
\begin{equation} \label{boson}
=\Upsilon_{bf} n_b \frac{\partial n_f}{\partial x}+\frac{1}{2}
\Upsilon_{2bf} n_b  \frac{{\partial}^3 n_f}{\partial x^3 } ,
\end{equation}
the continuity equation for fermions
\begin{equation}
\frac{\partial n_f}{\partial t}+\frac{\partial (n_f v_f
)}{\partial x}=0 ,
\end{equation}
and the momentum balance equation for fermions
$$m_f n_f  \frac{\partial v_f}{\partial t}+\frac{1}{2} m_f n_f  \frac{\partial v_f^2}{\partial x}-\frac{\hbar^2}{2m_f } n_f  \frac{\partial}{\partial x}  \biggl( \frac{ 1 }{\sqrt{n_f}}\frac{\partial^2 \sqrt{n_f}}{{\partial x}^2}\biggr)$$
$$+\frac{3}{8} \Upsilon_{2ff}  \frac{\partial n_f}{\partial x} \frac{\partial n_f^2}{\partial x}-\frac{3}{8} n_f \Upsilon_{2ff}  \frac{{\partial}^3 n_f}{{\partial x}^3}$$
$$ +\frac{1}{2} {(3\pi ^2)}^{2/3} \Upsilon_{2ff} \frac{\partial {n_f}^{8/3}}{\partial x} + \frac{\hbar ^2}{5 m_f}{(3\pi^2)}^{2/3} \frac{\partial {n_f}^{5/3}}{\partial x} $$
\begin{equation} \label{fermion}
=\Upsilon_{bf} n_f \frac{\partial n_b}{\partial x}+\frac{1}{2}
\Upsilon_{2bf} n_f  \frac{{\partial}^3 n_b}{\partial x^3 }.
\end{equation}
The equations given above utilize following designations: $m_b$,
$m_f$ - masses of boson and fermion atoms, respectively; $n_b$,
$n_f$ - concentrations of bosons and fermions; $v_b$, $v_f$ -
respective velocity fields. $\Upsilon_{bb}$, $\Upsilon_{bf}$ -
coefficients of boson-boson and boson-fermion interactions at the
first order of the interaction range. $\Upsilon_{2bb}$,
$\Upsilon_{2bf}$, $\Upsilon_{2ff}$ - coefficients of boson-boson,
boson-fermion and fermion-fermion interactions at the third order
of the interaction range. Interaction coefficients can be defined
with following equations:
\begin{equation} \label{Uij}
\Upsilon_{ij}=\frac{4\pi}{3} \int{dr r^3 \frac{\partial U_{ij}
(r)}{\partial r}},
\end{equation}
\begin{equation} \label{U2ij}
\Upsilon_{2ij}=\frac{4\pi}{15} \int{dr r^5 \frac{\partial U_{ij}
(r)}{\partial r}}.
\end{equation}

Presented here hydrodynamics equations correspond to the system of
two nonlinear Schrodinger equations one for bosons another for
fermions \cite{Andreev PRA08}. These nonlinear Schrodinger
equations are nonlocal and integro-differential. The nonlinear
Schrodinger equation for Bose particles is a generalization of
Gross-Pitaevskii equation. This generalization appear due to more
detailed account of short-range interaction up to third order on
interaction radius. For the first time a nonlocal Gross-Pitaevskii
equation were derived in Ref.s. ~\cite{Rosanov}, ~\cite{Braaten}.
This equation does not contain integral terms. Brief comparison
for different nonlocal generalization of the Gross-Pitaevskii
equation is discussed in ref. ~\cite{Andreev RPJ 11}.

Coefficients $\Upsilon_{bb}$ and $\Upsilon_{bf}$ are related to
interactions coefficients $g_{bb}$ and $g_{bf}$ of the
Gross-Pitaevskii equation as follows:
\begin{equation} \begin{array}{cc}
\Upsilon_{bb}=-g_{bb} ,& \Upsilon_{bf}=-g_{bf}.\end{array}
\end{equation}

It should be noted that the following relationship exists between
interaction coefficients and scattering amplitude $a_{ij}$
\begin{equation} \label{aij}
\Upsilon_{ij}=-\frac{4\pi \hbar^2 a_{ij}}{m}.
\end{equation}

All equations above are approximated up to the third order of the
interaction range. If terms containing $\Upsilon_{2bb}$,
$\Upsilon_{2bf}$, $\Upsilon_{2ff}$ , are neglected then the
resulting equation set corresponds to the first-order
approximation (the Gross-Pitaevskii approximation) which has been
used, for example, in \cite{Ramachandhran PRA 11}. It was shown in
\cite{Andreev PRA08} that the approximation is possible that
derives coefficients $\Upsilon_{2bb}$, $\Upsilon_{2bf}$ from
$\Upsilon_{bb}$, $\Upsilon_{bf}$. The explicit form of this
approximation is:
$$\Upsilon_{2bb} \simeq r_0^{-2} \Upsilon_{bb}$$
and
$$\Upsilon_{2bf} \simeq r_0^{-2} \Upsilon_{bf},$$
where $r_0$ - is a constant of the same order of magnitude as the atomic radius.\\

\section{\label{sec:level1} III. The soliton solution in a boson-fermion
mixture.}

The equation set in question should be expected to have soliton
solutions of a new type due to more accurate recognition of atomic
interactions. The method of perturbations may be applied to find
this soliton \cite{Washimi PRL 66, Kalita PlasmaPhys 98}.
According to this method all hydrodynamic values may be
represented as:

\begin{equation}
n_b=n_{0b}+\varepsilon n_{1b}+\varepsilon^2 n_{2b}+... ,
\end{equation}
\begin{equation}
n_f=n_{0f}+\varepsilon n_{1f}+\varepsilon^2 n_{2f}+... ,
\end{equation}
\begin{equation}
v_b=\varepsilon v_{1b}+\varepsilon^2 v_{2b}+... ,
\end{equation}
\begin{equation}
v_f=\varepsilon v_{1f}+\varepsilon^2 v_{2f}+... .
\end{equation}

We also performed the following "scaling" of variables:
\begin{equation}
\xi = \varepsilon ^{1/2}(x-Ut)
\end{equation}
and
\begin{equation}
\tau = \varepsilon ^{3/2}Ut .
\end{equation}
The latter expression introduces so-called "slow" time.

The following expression of the phase velocity $U$ can be derived
from equations (\ref{boson_nepr}-\ref{fermion}) in the first order
of the small parameter $\varepsilon$:
$$U_{\pm}^2=\frac{1}{2m_b m_f } \left(\Theta m_b n_{0f}-m_f n_{0b} \Upsilon_{bb} \right)$$
$$ \pm \frac{1}{2m_b m_f } \biggl ((\Theta m_b n_{0f}-m_f n_{0b} \Upsilon_{bb})^2
$$
\begin{equation} \label{U}
+4m_b m_f n_{0b} n_{0f} (\Theta\Upsilon_{bb}+\Upsilon_{bf}^2 )
\biggr)^{1/2} ,
\end{equation}
where \begin{equation} \Theta \equiv \frac{(3\pi^2
)^{2/3}}{n_{0f}} \left (\frac{\hbar^2}{3m_f }
n_{0f}^{2/3}+\frac{4}{3} \Upsilon_{2ff} n_{0f}^{5/3} \right).
\end{equation}
Quantity $\Theta$ describes the contribution of fermion dynamics
in boson-fermion mixture. $\Theta$ consists of two parts. The
first one appears from Fermi pressure and the second term presents
the fermion-fermion short-range interaction which arises at
interaction account up to the third order of the interaction
radius. In formula (\ref{U}) the quantity under radical is
positive.

Since our mixture is comprised by two interacting subsystems,
boson atoms and fermion atoms, two different values of phase
velocity are obtained here. We call phase velocity $U_{+}$ which
has "+" sign before the radical, the "fermion branch" of the
solution. Phase velocity $U_{-}$ is called the "boson branch" of
the solution by analogy. We introduced these terms to emphasize
the fact that in the absence of boson-fermion interactions
$\Upsilon_{bf}=0$ the system can be divided in two subsystems of
bosons and fermions, respectively, which do not interact. Phase
velocity $U_{+}$ corresponds in this case to fermions
$$U_{+}^{2}=\frac{n_{0f}}{m_{f}}\Theta,$$
and $U_{-}$ corresponds to bosons
$$U_{-}^{2}=-\frac{n_{0b}}{m_{b}}\Upsilon_{bb}.$$
We can see that $U_{-}^{2}>0$ for particles with repulsive
interaction $\Upsilon_{bb}<0$.

In the second order of the small parameter $\varepsilon$ we derive
the Korteweg - de Vries equation for small perturbations of boson
concentration in a boson-fermion mixture:
\begin{equation} \label{KdF}
p \frac{\partial n_{1b}}{\partial \tau}+q \frac{\partial^3
n_{1b}}{\partial \xi^3 }+s n_{1b}  \frac{\partial n_{1b}}{\partial
\xi}=0,
\end{equation}
where coefficient  $p$ of the term containing slow time
\begin{equation} \label{p0}
p=2 U^2 (\Theta m_b n_{0f} - \Upsilon_{bb} n_{0b} m_f - 2U^2 m_b
m_f),
\end{equation}
coefficient $q$ of the term that corresponds to dispersion
$$q=-(\Theta n_{0f}-m_f U^2)\times$$
$$ \times\left (\frac{\hbar^2}{4m_f}+ \frac{1}{8}\Upsilon_{2bb}n_{0b}-\frac{1}{2}\Upsilon_{2bf}n_{0b}\frac{\Upsilon_{bb}}{\Upsilon_{bf}}-\frac{1}{2}\Upsilon_{2bf}\frac{U^2 m_b}{\Upsilon_{bf}} \right)$$
$$+ \biggl(\frac{\hbar^2}{4m_f}+\frac{1}{4}\Upsilon_{2ff}n_{0f}\biggr)(n_{0b}\Upsilon_{bb}+U^2 m_{b})$$
\begin{equation}
  +
\frac{1}{2}\Upsilon_{2bf}\Upsilon_{bf}n_{0b}n_{0f},
\end{equation}
and coefficient $s$ of the non-linear term
$$ s=3\frac{U^2m_b}{n_{0b}} (\Theta n_{0f}-U^2 m_f)+\Upsilon_{bf}(\Upsilon_{bb} n_{0b}+U^2 m_b) $$
$$+\left (\Upsilon_{bb}+\frac{U^2 m_b}{n_{0b}} \right)^2
\frac{n_{0b}}{\Upsilon_{bb}}\times$$
\begin{equation}
\times\left (2\frac{U^2 m_f}{n_{0f}}+\frac{20}{9}(3\pi^2
n_{0f})^{2/3} \Upsilon_{2ff}+\frac{\hbar^2}{9 m_f}
\frac{(3\pi^2)^{2/3}}{n_{0f}^{1/3}} \right).
\end{equation}

The soliton solution of the Korteweg - de Vries equation
(\ref{KdF}) is well known in the form of
\begin{equation} \label{soliton}
n_{1b}=\frac{3pV}{s} \cdot \frac{1}{\cosh^2
\left(\sqrt{\frac{Vp}{4q}}\eta \right )},
\end{equation}
where $\eta=\xi-V\tau.$

Below we discuss perturbations in bosons implying that
perturbations in fermions are similar to them due to the following
linear relationship between $n_{1b}$ and $n_{1f}$:
\begin{equation} \label{svyaz_boson_fermion}
\Upsilon_{bf} n_{1f}=- \left (\Upsilon_{bb}+\frac{U^2
m_b}{n_{0b}} \right ) n_{1b}.
\end{equation}

Soliton pairs (in bosons and fermions, respectively) of two types
may evolve: dark-bright and dark-dark.

\section{\label{sec:level1} IV. The fermion branch of the solution.}

Let's investigate "fermion branch" of the solution, where phase
velocity $U$ from the equation  (\ref{U}) is taken with $"+"$
sign.

If boson-fermion interactions are "strong" enough compared to
boson-boson interactions, then $ U^2>0 $. We use apply assumption
to the fermion branch below.

For numerical analysis of our results we introduce following
dimensionless variables: dimensionless phase velocity
$W_{+}=m_{f}U_{+}/(\hbar n_{0f}^{1/3})$, mass rate $\mu=m_{f}/m_{b}$,
concentrations rate $N=\sqrt{n_{0b}/n_{0f}}$, and dimensionless
interaction constants
$\gamma_{bb}=m_{f}n_{0b}\Upsilon_{bb}/(n_{0f}^{1/3}\hbar)^{2}$,
$\gamma_{bf}=m_{f}\sqrt{n_{0b}n_{0f}}\Upsilon_{bf}/(n_{0f}^{1/3}\hbar)^{2}$,
$\gamma_{ff}=m_{f}n_{0f}\Upsilon_{2ff}/\hbar^{2}$,
$B_{bb}=m_{f}n_{0b}\Upsilon_{2bb}/\hbar^{2}$, and
$B_{bf}=n_{of}^{2/3}\Upsilon_{2bf}/\Upsilon_{bf}$.

Phase velocity for fermion branch is shown at Fig. 1.

It can be shown that
\begin{equation} \label{p} p<0 \end{equation}
at any allowable values of physical parameters. It follows from
numerical analysis of $p$.

\begin{figure}
\includegraphics[width=8cm, angle=0]{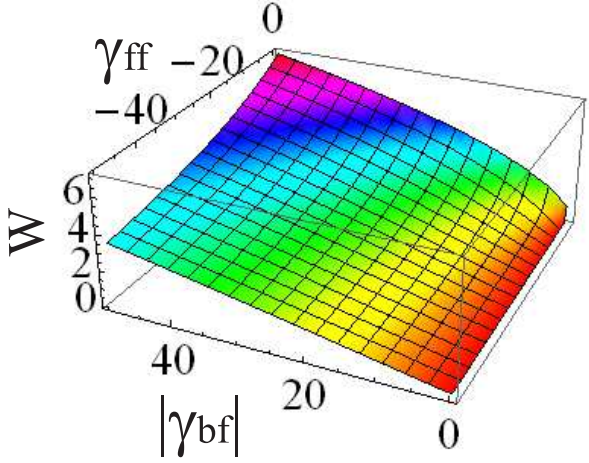}
\caption{\label{fig1:epsart} (Color online) The relationship
between dimensionless phase velocity $W_{f}$ and coefficients of
fermion-fermion interaction $\gamma_{bf}$ and boson-fermion
interaction $\gamma_{bf}$ for fermion branch. Figure is made for the following parameters of the system: $\gamma_{bb}=-10^{-2}$, $\mu=1$, $N=1$, $B_{bf}=10^{-3}$, and $B_{bb}=-10^{-4}$. This plot and plots 2-4 made for the repulsion between Bose particles $\gamma_{bb}<0$. Dimensionless phase velocity $W_{f}$ has analogous form in the case of attraction between Bose particles.}
\end{figure}

As a starting point we considered the first order of the
interaction range, neglecting the contribution of terms that are
proportional to coefficients $\Upsilon_{2bb}$, $\Upsilon_{2bf}$,
$\Upsilon_{2ff}$ in equations (\ref{boson}) and (\ref{fermion}).

In this case the coefficient $q$ takes the following value:
\begin{equation}
\label{q} q= \alpha \frac{\hbar^2}{4 m_b} + \beta \frac{\hbar^2}{4
m_f}>0,
\end{equation}
where $\alpha>0$ and $\beta>0$. Evident form for $\alpha$ and
$\beta$ to be
$$\alpha=\frac{1}{2m_b}  \biggl(-(\Theta m_b n_{0f}+\Upsilon_{bb} m_f n_{0b})$$
 \begin{equation}
+\sqrt{ (\Theta m_b n_{0f}+\Upsilon_{bb} m_f n_{0b})^2 + 4 m_b m_f
n_{0b} n_{0f} (\Upsilon_{bf})^2}  \biggr),
\end{equation}
and
$$\beta=\frac{1}{2m_f} \biggl((\Theta m_b n_{0f}+\Upsilon_{bb} m_f n_{0b})$$
\begin{equation}
+\sqrt{(\Theta m_b n_{0f}+\Upsilon_{bb} m_f n_{0b})^2 + 4 m_b m_f
n_{0b} n_{0f} (\Upsilon_{bf})^2} \biggr).
\end{equation}

Expressions (\ref{soliton}), (\ref{p}) and (\ref{q}) make it clear
that no soliton exists in the boson-fermion mixture in the
Gross-Pitaevskii approximation, as the radicand in (\ref{soliton})
is negative.

So, we took into account the third order by the interaction radius.

Now the coefficient  $q$ has the following form:
\begin{equation} \label{q0}
q=\beta \left (\frac{\alpha}{\beta}
\frac{\hbar^2}{4m_b}+\frac{\hbar^2}{4m_f} +\frac{1}{4} n_{0f}
Y_{2ff} \right )+...
\end{equation}
It means that if the repulsion of fermions is strong enough (i.e.
the coefficient of fermion-fermion interaction $Y_{2ff}<0$) the
$q$ value would be negative. Strong fermion-fermion interactions
in the boson-fermion mixture make possible perturbations of a new
type, which do not occur in the first order of the interaction
range.

\begin{figure}
\includegraphics[width=8cm, angle=0]{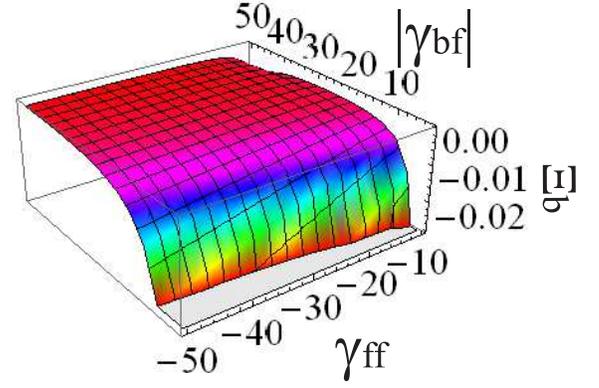}
\caption{\label{fig1:epsart} (Color online) The relationship
between soliton amplitude $\Xi_{b}$ of Bose subsystem \emph{and} the coefficient of
boson-fermion interaction $\gamma_{bf}$ and fermion-fermion interaction $\gamma_{ff}$ for fermion branch.
Figure is made for the following parameters of the system: $\gamma_{bb}=-10^{-2}$, $\mu=1$, $N=1$, $B_{bf}=10^{-3}$, and $B_{bb}=-10^{-4}$.}
\end{figure}

In the case the considered perturbations of concentration are
small, solitons of two kinds may arise: solitons of fermion
rarefication and of fermion compression. The soliton type is
determined by the sign of its dimensionless amplitude of bosons
\begin{equation}
\Xi_{b}=\frac{3pV}{s n_{0b}},
\end{equation}
and fermions $\Xi_{f}$ which connected with the $\Xi_{b}$ by formula (\ref{svyaz_boson_fermion}).

The dependence between amplitude of soliton in the boson subsystem
and the coefficient of boson-fermion interaction  $\Upsilon_{bf}$
is shown at Fig.s 2-4. If the coefficient value is high positive (it
corresponds to a strong attraction between boson and fermion
atoms) then boson perturbations would be solitons of depression (dark soliton, it is shown at Fig.2)
and, as it follows from (\ref{svyaz_boson_fermion}), fermion
perturbations would be solitons of compression (bright soliton) for $\gamma_{bf}>0$ or soliton of depression (dark soliton) for $\gamma_{bf}<0$ (it is shown at Fig.s 3 and 4). These are
dark-bright and dark-dark solitons in a two-component system of bosons and
fermions. So, a strong repulsion between bosons and fermions (i.e.
negative values of the  $\Upsilon_{bf}$ coefficient) would lead to
solitons of compression of bosons and fermions, which
correspond to a dark-dark soliton.

The dimensionless soliton width $D$ can be defined by the
following expression:
\begin{equation} \label{D}
D=\sqrt{\frac{4qn_{0f}^{2/3}}{Vp}}
\end{equation}

It follows from  (\ref{p0}) and (\ref{q0}) that soliton width
depends on the fermion-fermion interaction. This dependence is
plotted at Fig. 5. Note that the surface's shape does not
change along with the sign of $\gamma_{bf}$ as the expression for
the soliton width $D$ contains the coefficient of boson-fermion
interaction in a squared form.

\begin{figure}
\includegraphics[width=8cm, angle=0]{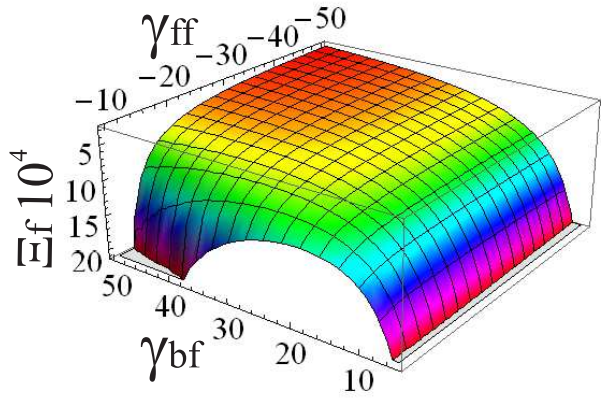}
\caption{\label{fig1:epsart} (Color online) The dependence curve
of the soliton amplitude $\Xi_{f}$ of Fermi subsystem vs the coefficient of fermion-fermion
interaction $\gamma_{ff}$ and boson-fermion interaction $\gamma_{bf}$ for the fermion
branch. The plot is built with following parameters: $\gamma_{bb}=-10^{-2}$, $\mu=1$, $N=1$, $B_{bf}=10^{-3}$, and $B_{bb}=-10^{-4}$.}
\end{figure}

\begin{figure}
\includegraphics[width=8cm, angle=0]{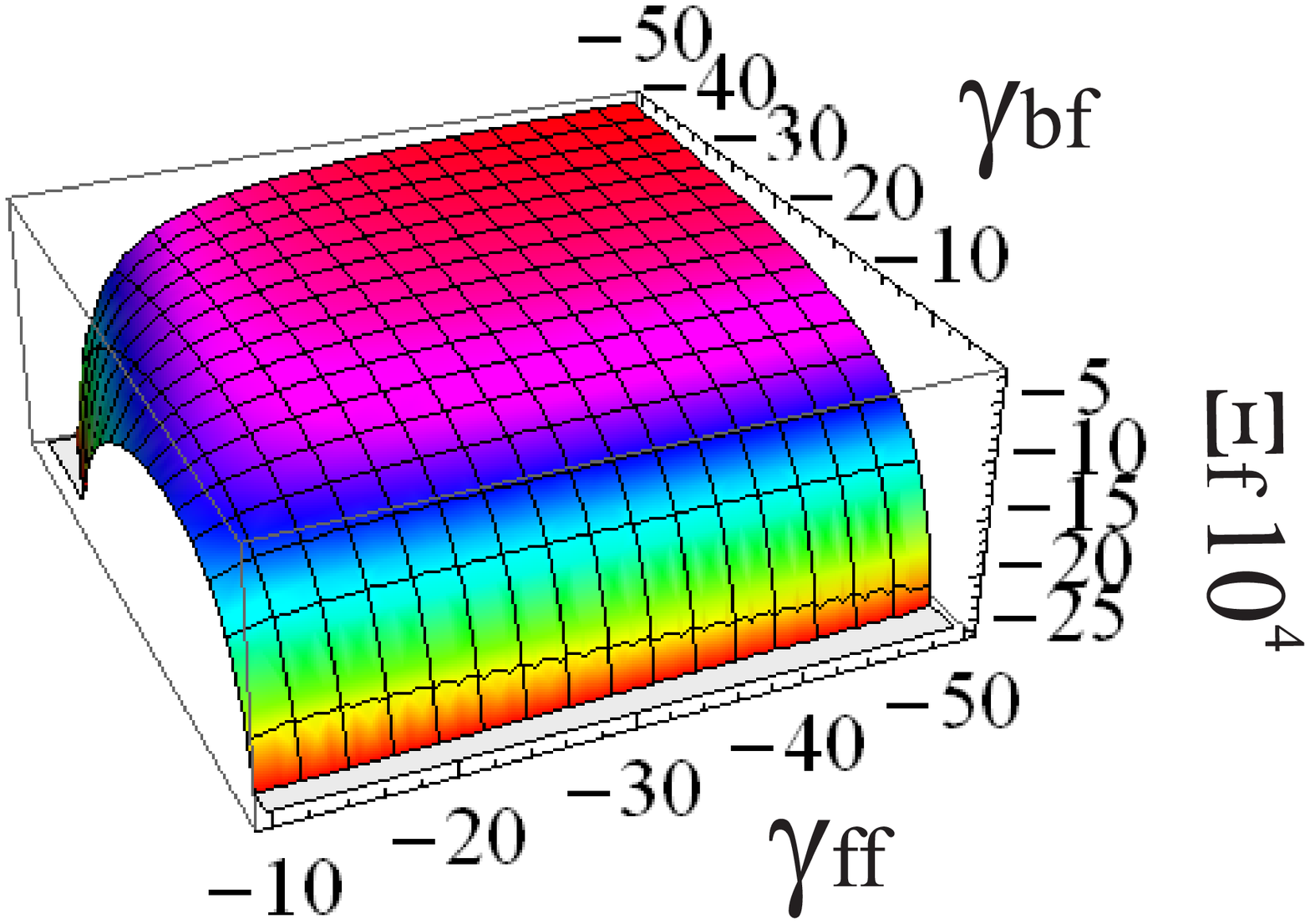}
\caption{\label{fig1:epsart} (Color online) The dependence curve
of the soliton amplitude $\Xi_{f}$ of Fermi subsystem vs the coefficient of fermion-fermion
interaction $\gamma_{ff}$ and boson-fermion interaction $\gamma_{bf}$ for the fermion
branch in the case negative $\gamma_{bf}$ that corresponds to repulsion between Bose and Fermi particle. The plot is built with following parameters: $\gamma_{bb}=-10^{-2}$, $\mu=1$, $N=1$, $B_{bf}=10^{-3}$, and $B_{bb}=-10^{-4}$.}
\end{figure}

\begin{figure}
\includegraphics[width=8cm, angle=0]{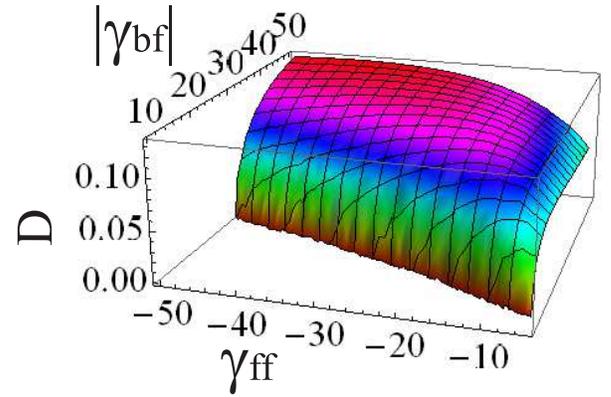}
\caption{\label{fig1:epsart} (Color online) The dependence curve
of the soliton width $D$ vs the coefficient of fermion-fermion
interactions $\gamma_{ff}$ and the coefficient of boson-fermion
interaction $\gamma_{bf}$ for the fermion branch expressed in
reduced values. The plot is built with following parameters:
$\gamma_{bb}=-10^{-2}$, $\mu=1$, $N=1$, $B_{bf}=10^{-3}$, and
$B_{bb}=-10^{-4}$. Analogous dependence take place at
$\gamma_{bb}>0$ for $\gamma_{bb}\sim10^{-2}\div10^{-3}$ in the
area of large boson fermion interaction $|\gamma_{bf}| >10$.}
\end{figure}

There is interesting behavior of this solution for attractive
boson-boson interaction. In this case soliton exist in two areas
of system parameters. For the large $|\gamma_{bf}|$
($|\gamma_{bf}|>10$), function $D(\gamma_{ff}, \gamma_{bf})$ is
analogous to the same presented at Fig. 5. But for $\gamma_{bb}>0$
there is an area at the small $|\gamma_{bf}|$ where exist soliton
solution. In this area soliton width is large, much more than 1.
It is presented at Fig.s 6-8.

In the case $\gamma_{bb}>0$ and $|\gamma_{bf}|<1$ amplitude of bosons is positive $\Xi_{b}>0$. So, we have bright soliton in subsystem of bosons Fig. 6. From Fig. 7 we can see that if $\gamma_{bf}<0$ amplitude of soliton in Fermi subsystem is positive. Thus, we have found bright soliton in Fermi subsystem. If $\gamma_{bf}>0$ we have dark soliton in Fermi subsystem. In the result in boson-fermion mixture we have bright-dark soliton for $\gamma_{bf}<0$, and bright-bright soliton for $\gamma_{bf}>0$.

Considered in the paper values of interaction parameters may be
obtained using Feshbach resonance \cite{Cheng Chin RMP 10, Bloch
RMP 08}. It follows from (\ref{Uij}) and (\ref{U2ij}), that
changing potential of atomic interaction changes both
$\Upsilon_{ij}$ and $\Upsilon_{2ij}$ coefficients. Note that
soliton of this type occurs due to boson-fermion interactions.

\begin{figure}
\includegraphics[width=8cm, angle=0]{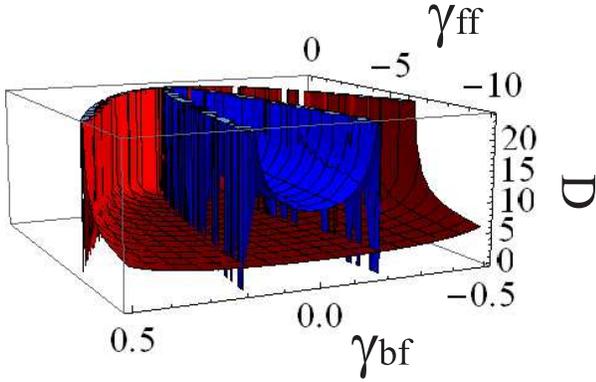}
\caption{\label{fig5:epsart} (Color online) The dependence curve
of the soliton width  $D$ the coefficient of boson-fermion
$\gamma_{bf}$ and fermion-fermion $\gamma_{ff}$ interaction for
the fermion branch. This plot made for the attraction between Bose
particles $\gamma>0$. The plot is built for two values of
$\gamma_{bb}$ (for blue-narrow surface we took
$\gamma_{bb}=10^{-3}$, and for red-wide surface we used
$\gamma_{bb}=10^{-2}$) with following parameters: $\mu=1$, $N=1$,
$B_{bf}=10^{-4}$, and $B_{bb}=-10^{-4}$.}
\end{figure}

\begin{figure}
\includegraphics[width=8cm, angle=0]{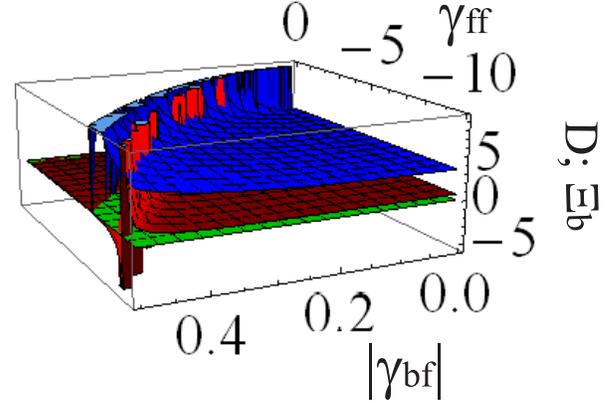}
\caption{\label{fig6:epsart} (Color online) The dependence curve
of the soliton width  $D$ (blue-top surface) and amplitude for
bosons $\Xi_{b}$ (red-middle surface) on the coefficient of
boson-fermion $\gamma_{bf}$ and fermion-fermion $\gamma_{ff}$
interaction for the fermion branch. Soliton exist if its width $D$
is positive. We present soliton width  $D$ and amplitude for
bosons $\Xi_{b}$ on the same plot to show the value of amplitude
in the area where soliton exist. Green plane-lower surface present
zero level. It show us that amplitude is positive. The plot is
built with following parameters: $\gamma_{bb}=10^{-2}$, $\mu=1$,
$N=1$, $B_{bf}=10^{-4}$, and $B_{bb}=-10^{-4}$.}
\end{figure}

\begin{figure}
\includegraphics[width=8cm, angle=0]{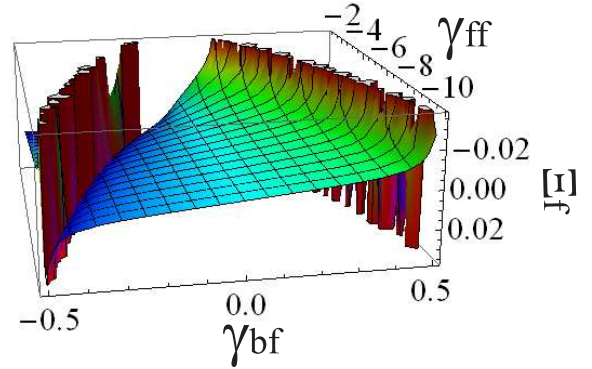}
\caption{\label{fig7:epsart} (Color online) The dependence curve
of the soliton width  $D$ the coefficient of boson-fermion
$\gamma_{bf}$ and fermion-fermion $\gamma_{ff}$ interaction for
the fermion branch. The plot is built with following parameters:
$\gamma_{bb}=10^{-2}$, $\mu=1$, $N=1$, $B_{bf}=10^{-4}$, and
$B_{bb}=-10^{-4}$.}
\end{figure}

\section{\label{sec:level1} V. The boson branch of the solution.}

To obtain the boson branch phase velocity $U$ from the equation
(\ref{U}) should be taken with "$-$" sign.

Using the first order of the interaction radius does not allow to
predict solitonic perturbations in boson-fermion mixtures.
It can be done only when the third order is taken into account. If
the mixture of bosons and fermions divides into two subsystems
that do not interact, bosons "do not feel" the presence of
fermions, $\Upsilon_{bf}=0$, and the coefficient $q \simeq -
\hbar^2/4 m_b - n_{0b}\Upsilon_{2bb}/8 $ ~\cite{Andreev arxiv 11
1}. This means that repulsion of bosons ($\Upsilon_{2bb}<0$) must
be strong enough to cause soliton formation in a subsystem of
boson atoms.

If the interaction of bosons and fermions occurs, then the
acceptable range of physical values, where the soliton formation is
possible, changes. In boson-fermion mixture a soliton corresponding
to boson branch exists at more large strength of boson-boson
interaction, in comparison with the case when $\Upsilon_{bf}=0$.
We have bright soliton solution ($\Xi_{b}>0$) in Bose subsystem.
In Fermi subsystem we find dark $\Xi_{f}<0$  (bright $\Xi_{f}>0$)
soliton for repulsion (attraction) between Bose and Fermi
particles $\Upsilon_{bf}<0$ ($\Upsilon_{bf}>0$). We have got it
because quantity $-(\Upsilon_{bb}+m_{b}U_{-}^{2}/n_{0b})$
presented by formula (\ref{svyaz_boson_fermion}) is positive.
Consequently, $\Xi_{f}$ has the same sign as $\Upsilon_{bf}$. So,
in a mixture of bosons and fermions this type of soliton solution
may exist.

Finally, we can note that two soliton solutions exists in
boson-fermion mixture due to account of short range interaction up
to the third order of the interaction radius.

\section{\label{sec:level1} VI. Conclusion}

The quantum hydrodynamics approach is well suitable to derive
equations that describe multiparticle quantum systems. In our
case we applied the QHD approach to the ultra-cold mixture of
boson and fermion atoms with short-range interaction potential
in order to analyze the possibility of small-scale solitonic
perturbations. We show that solitons of two novel types may
occur, which are related to fermion and boson branches of the
solution, respectively.

In the fermion branch the soliton of a new type occurs due to
boson-fermion interactions at the first order of the interaction
range. Fermion-fermion interactions also play substantial role in
formation of these perturbations. The type of soliton formed in
the system (compressed or rarefied) is determined by the sign of
boson-fermion interaction, i.e.  repulsion or attraction,
respectively. The dependence of fermion concentration on system
characteristics was successfully derived. Consideration of the
third order of the interaction range is the key factor that
allowed prediction of a novel type of soliton solution. It also
helped us to find new acceptable ranges of physical values, where
the existence of soliton is possible for the boson branch of the
solution in the boson-fermion mixture, compared to the system
comprised of bosons only.

\end{document}